%% file: main.tex
\newif\ifDEBUG
\DEBUGfalse

\newif\ifANONYMOUS
\ANONYMOUSfalse

\newif\ifARXIV
\ARXIVtrue

\ifANONYMOUS
  \documentclass[sigconf,review,anonymous=false,natbib=false]{acmart}
\else
  \ifARXIV
    \documentclass[sigconf,anonymous=false,natbib=false]{acmart}
  \else
    \documentclass[sigconf,anonymous=false,natbib=false]{acmart}
  \fi
\fi

\AtBeginDocument{%
  }

\setcopyright{cc}
\copyrightyear{2026}
\acmYear{2026}
\acmDOI{}
\acmISBN{}
\acmConference[JAWs 2026]{Journal Ahead Workshop (JAWs) 2026}{April 2026}{Rio de Janeiro, Brazil}
\acmBooktitle{Journal Ahead Workshop (JAWs) 2026}

\RequirePackage[
  datamodel=acmdatamodel,
  style=acmnumeric,
  ]{biblatex}

\input{typesetting}

\begin{document}

\title[\toolname]{ {\toolname}: Systematic Testing of Regular Expression Engines}
\title{Towards the Systematic Testing of Regular Expression Engines}

\author{Berk Çakar}
\orcid{0009-0006-6613-5591}
\affiliation{%
  \institution{Electrical and Computer Engineering\\Purdue University}
  \city{West Lafayette}
  \state{IN}
  \country{USA}}
\email{bcakar@purdue.edu}

\author{Dongyoon Lee}
\orcid{0000-0002-2240-3316}
\affiliation{%
  \institution{Computer Science\\Stony Brook University}
  \city{Stony Brook}
  \state{NY}
  \country{USA}
}
\email{dongyoon@cs.stonybrook.edu}

\author{James C. Davis}
\orcid{0000-0003-2495-686X}
\affiliation{%
  \institution{Electrical and Computer Engineering\\Purdue University}
  \city{West Lafayette}
  \state{IN}
  \country{USA}
}
\email{davisjam@purdue.edu}

\authorsaddresses{%
  Authors' Contact Information: \href{https://orcid.org/0009-0006-6613-5591}{Berk Çakar}, bcakar@purdue.edu;
  \href{https://orcid.org/0000-0002-2240-3316}{Dongyoon Lee}, dongyoon@cs.stonybrook.edu;
  \href{https://orcid.org/0000-0003-2495-686X}{James C. Davis}, davisjam@purdue.edu.%
}

\renewcommand{\shortauthors}{Çakar, Lee, and Davis}

\begin{abstract}
    Software engineers use regular expressions (regexes) across a wide range of domains and tasks.
    To support regexes, software projects must integrate a regex engine, whether provided natively by the language runtime (\eg Python's re) or included as an external dependency (\eg PCRE).
    However, these engines may contain bugs and introduce vulnerabilities.
    A common strategy for testing regex engines involves differential testing---comparing outputs across different implementations.
    However, this approach is concerning because regex syntax and semantics vary significantly between dialects (\eg POSIX \vs PCRE).
    Fuzzing is also utilized to ease testing of feature-rich regex implementations to expose defects, but naive byte-level mutations generate syntactically invalid inputs that exercise only parsing logic, not matching internals.

    In this work, we describe our progress towards \toolname, a framework that systematically tests regular expression engines by combining grammar-aware fuzzing for high code coverage with metamorphic testing to generate dialect-independent test oracles.
    So far, we have surveyed testing practices across 22 regex engines, analyzed 1,007 regex engine bugs and 156 CVEs to characterize failure modes, and curated 16 metamorphic relations for regexes derived from Kleene algebra.
    Our preliminary evaluation on PCRE shows that \toolname achieves $3\times$ higher edge coverage than existing fuzzing approaches and has identified three new memory safety defects.
    We conclude by describing our next steps toward our ultimate goal: helping regex engine developers identify bugs without depending on a consistent cross-implementation standard.
\end{abstract}



\maketitle

\section{Introduction}\label{sec:introduction}

In this paper, we are working towards improving the testing techniques for a critical class of software infrastructure: regular expression (regex) engines.
Regexes are a fundamental tool in software engineering, used extensively for pattern matching~\cite{friedlMasteringRegularExpressions2002}, input validation~\cite{10.5555/2337223.2337334,barlasExploitingInputSanitization2022}, text processing~\cite{gnu_emacs_regexps,msft_visualstudio_regex}, and data extraction~\cite{7374717,10.1145/1806799.1806821}.
In software projects, regexes are supported through regex engines---software components that interpret regex patterns and execute matching operations against input strings.
These engines are either built into language runtimes (\eg Python's re module~\cite{pythonre}, JavaScript's (V8) Irregexp~\cite{v8regexp}) or provided as external libraries (\eg PCRE~\cite{pcre2024}, RE2~\cite{re2}, Oniguruma~\cite{oniguruma}).
However, regex engines are not immune to bugs.
One of the most widely used regex engines, PCRE2, has had 10+ CVEs (all rated high or critical) and 140+ confirmed bug reports since 2015 (\cref{sec:bugs_vulns_in_engines}).

Current testing practices for regular expression engines are largely \adhoc and inconsistent.
Our analysis of prominent regex engine implementations (\cref{subsec:regex_engines}) shows that most engines rely primarily on unit and regression testing, which demand substantial manual effort.
Differential testing is widely used to identify semantic discrepancies, but its effectiveness is limited by the lack of a universal regex standard: engines implement distinct dialects with subtle semantic differences, causing differential testing to generate false positives that reflect expected dialect variation rather than genuine defects (\cref{sec:challenge}).
Fuzzing is also utilized for the discovery of safety violations, yet existing approaches are typically naive byte-level mutations that offer limited coverage of matching logic.
Ultimately, existing testing practices suffer from the absence of both a unified, systematic testing methodology and a common semantic oracle.

In this paper, we present our work towards \toolname, a systematic testing framework that combines grammar-aware fuzzing with metamorphic testing.
To automate the discovery of code paths, \toolname employs a grammar-aware mutation strategy: it parses regexes into abstract syntax trees (ASTs) and performs structured subtree replacements seeded from a large-scale corpus of real-world patterns~\cite{cakar2025reuseneedsystematiccomparison}.
Unlike naive byte-level fuzzing---which is likely to generate invalid inputs that are rejected during parsing---this structure-aware approach ensures generated patterns are syntactically valid, thereby exercising the engine's matching logic~\cite{aschermann2019nautilus,fuzzingbook2024:Grammars}.
While fuzzing with runtime sanitizers~\cite{ubsan,Serebryany2012AddressSanitizer,Stepanov2015MemorySanitizer} effectively identifies safety violations (\eg buffer overflows, use-after-free errors, integer overflows, and undefined behavior), it cannot detect non-crashing semantic errors.
Therefore, \toolname supplements crash oracles with metamorphic testing~\cite{10.1145/3143561,Segura2016Metamorphic}: by verifying algebraic properties (\eg $r^{**} \equiv r^{*}$) within a single engine, \toolname provides a semantic oracle without the false positives in differential testing.

\pagebreak

Our contributions so far are:

\begin{itemize}
    \item \textbf{Empirical Study of Testing Practices.} We survey 22 regex engines and characterize 12 testing techniques employed in practice, revealing that 82\% of bugs are user-reported and only 2 of 22 engines employ grammar-aware fuzzing (\cref{sec:testing_practices}).

    \item \textbf{Bug and CVE Analysis.} We analyze 1,007 bugs across 11 third-party engines and 156 CVEs across the regex engine landscape, finding that semantic bugs (35\%) and memory safety vulnerabilities (52\% of CVEs) dominate (\cref{sec:bugs_vulns_in_engines}).

    \item \textbf{\toolname Framework.} We design and prototype \toolname, combining grammar-aware fuzzing with metamorphic testing (\cref{sec:design}).

    \item \textbf{Metamorphic Relations for Regexes.} We compile a catalog of 16 metamorphic relations, providing dialect-independent oracles for semantic bug detection (\cref{sec:design}).

    \item \textbf{Preliminary Evaluation.} We demonstrate \toolname prototype's feasibility on PCRE, achieving 3$\times$ higher coverage than baselines and discovering three memory safety bugs (\cref{sec:preliminary_evaluation}).
\end{itemize}

\vspace{0.1cm}
\noindent
In~\cref{sec:JournalNext}, we describe our planned future work.

\begin{table*}[ht]
\centering
\begin{threeparttable}
\caption{
Feature comparison of 22 regex engines based on commonly referred regex features and constructs.
The features and constructs are compiled from~\cite{davisImpactDefeatRegular2020,regular_expressions_info_flavors,pcre_manual,opengroup_posix_regex_ch9}.
For brevity, we inspect only the \textit{existence} of advanced features (\eg callouts) rather than specific syntax.
Notation: \textit{BT}=Backtracking, \textit{Auto}=Automata, \textit{Hyb}=Hybrid.
}
\label{tab:engine-comparison}
\tiny
\setlength{\tabcolsep}{0.8pt}
\input{tables/table_1}
\end{threeparttable}
\end{table*}

\section{Background and Related Work}\label{sec:background}

We introduce regex engines (\cref{subsec:regex_engines}), prior research on regex engine testing (\cref{subsec:regex_testing_research})%
, and metamorphic testing (\cref{subsec:metamorphic_testing}).

\subsection{Regex Engines}\label{subsec:regex_engines}

A \textit{regex engine} is a software component that compiles a regex pattern $r$ into an intermediate representation (\eg an automaton) and applies it to determine whether an input string $s$ matches $r$.
Most regex engines employ one of three algorithms~\cite{10.1145/3708821.3733912}\footnote{Microsoft's C\# has explored~\cite{10.1145/3591262,Saarikivi_Veanes_Wan_Xu_2019} Brzozowski derivatives~\cite{Brzozowski1964RegexDerivatives} as a fourth approach.}: (1)~\textit{backtracking}, which simulates an NFA via recursive exploration with exponential worst-case complexity but supports expressive features like backreferences and lookarounds; (2)~\textit{automata-based}, which uses Thompson's lockstep algorithm~\cite{thompson1968} for linear-time matching guarantees but cannot support expressive (\ie non-regular) features; or (3)~\textit{hybrid}, combines both strategies, employing au\-to\-ma\-ta-\-based matching wherever possible and falling back to backtracking for patterns requiring advanced features.

Formally, regex patterns restricted to concatenation, alternation, and Kleene star are called \textit{K-regexes}~\cite{Kleene1951NerveNetsAndRegularLanguages} and correspond to regular languages; patterns with additional features like backreferences, lookarounds, and atomic groups are called \textit{E-regexes} (extended regexes) and may recognize non-regular languages.~\cite{Campeanu_Salomaa_Yu_2003}.

Despite this theoretical foundation, no unified specification governs regex syntax and semantics.
Each engine implements its own \textit{dialect} (or \textit{flavor})~\cite{regular_expressions_info_flavors} of regexes.
Most dialects draw from two influential sources: the \textit{POSIX} standard~\cite{opengroup_posix_regex_ch9} and \textit{PCRE}~\cite{pcre2024} (Perl-Compatible Regular Expressions).
These sources differ in both constructs (\eg PCRE supports lookahead assertions and non-capturing groups absent from POSIX) and match semantics (\eg POSIX mandates leftmost-longest matching, whereas PCRE follows leftmost-first semantics based on alternative ordering).
Engines may adopt constructs from one, both, or a selective combination of these specifications, and may even introduce their own.
This lack of standardization results in non-portable regular expressions: the same pattern may exhibit different behavior across engines.
As Davis \etal~\cite{davis2019testing} observed, even among regexes that compile successfully in multiple languages, a significant fraction produce divergent results.
Regexes are thus far from being a \textit{lingua franca}~\cite{linguaFranca}---practitioners cannot assume that a pattern tested in one environment will behave identically elsewhere.

\vspace{-1mm}
\subsection{Research on Regex Engine Testing}\label{subsec:regex_testing_research}

Prior work on regular expressions focuses predominantly on testing patterns, not engines.
Research on test string generation~\cite{larsonGeneratingEvilTest2016,5431742}, ReDoS detection~\cite{davis_impact_2018,liReDoSHunterCombinedStatic2021,281448}, and pattern synthesis/repair~\cite{ liTransRegexMultimodalRegular2021,panAutomaticRepairRegular2019} all assume engine correctness while validating developer-written patterns.
Empirical studies of regex bugs likewise examine pattern-level faults rather than implementation defects~\cite{eghbaliNoStringsAttached2020,wangEmpiricalStudyRegular2020,wang_demystifying_2022}.

More broadly, testing language implementations and DSL engines offers transferable techniques.
Csmith's~\cite{10.1145/1993498.1993532} random program generation found 325+ compiler bugs, while grammar-based fuzzing has proven effective for language implementations (\eg~\cite{aschermann2019nautilus,180229,287210}).
For DBMS, Rigger and Su~\cite{riggerDetectingOptimizationBugs2020,riggerFindingBugsDatabase2020,riggerTestingDatabaseEngines2020} used metamorphic oracles to detect logic bugs. %
We portray regex engine testing in practice in \cref{sec:testing_practices}.

\subsection{Metamorphic Testing}\label{subsec:metamorphic_testing}

Metamorphic testing (MT) addresses the oracle problem—where expected outputs are unknown, by verifying relationships between multiple executions rather than individual results~\cite{chen1998metamorphic}. Given a source input and its output, metamorphic relations (MRs) define how transformed follow-up inputs should relate to the original output. MT has proven effective for testing compilers and language implementations: Le \etals Equivalence Modulo Inputs (EMI)~\cite{10.1145/2594291.2594334} detected 147 bugs in GCC and LLVM.
For DSLs specifically, Mansur \etal~\cite{10.1145/3468264.3468573} applied MT to Datalog engines, finding 13 query bugs.

In the context of regexes, an MR specifies a transformation $T$ where $T(r) \equiv r$, asserting that $\textsc{Match}(r, s) \sim \textsc{Match}(T(r), s)$ for all inputs $s$, where $\sim$ denotes the relation checked.

\section{Research Overview}\label{sec:research_overview}
Building the first testing framework for regex engines requires a comprehensive understanding of both the current of state of practice and the nature of defects in these systems. We structure our research around two phases and three questions that progress from empirical investigation to system design and evaluation.

\myparagraph{Phase 1: Empirical Investigation}
Existing literature provides little insight into how regex engines are validated in the wild or what types of bugs persist despite these efforts. To inform the design of systematic testing framework, we first establish a baseline of current practices and failure modes.

\begin{researchquestions}
    \item \textit{How are regular expression engines tested in practice?}\\We survey regex engines to determine the prevalence of testing techniques. This exploratory question sheds light on engineering practices across leading representatives of this class of software.
    \item \textit{What are the failure modes and their causes in regex engines?}\\By analyzing historical bug reports and CVEs, we learn whether existing testing gaps lead to reliability or security issues (\eg semantic errors \vs memory safety violations). %
\end{researchquestions}

\myparagraph{Phase 2: Design and Evaluation of the \toolname Framework} Based on the findings from RQ1 and RQ2, we propose \toolname. This framework is designed to close these gaps by combining \textit{grammar-aware fuzzing} (to address coverage limitations) with \textit{metamorphic testing} (to provide a dialect-independent semantic oracle).

\DY{For this, I wonder whether combining grammar-aware fuzzing and metamorphic testing yields strong synergy, or whether they are largely independent --- grammar-aware fuzzing  improves coverage, and metamorphic testing targets non-crash semantic bugs. }
\begin{researchquestions}[start=3]
    \item \textit{How can we achieve a systematic, dialect-independent testing framework for regex engines?} We assess the design and efficacy of \toolname through three sub-questions:
    \begin{researchquestions}
        \item \textit{Can we define a catalog of metamorphic relations that serve as effective semantic oracles?} %
        \item \textit{How does \toolname compare to other baselines?} %
        \item \textit{Is the framework capable of replicating previous bugs and discovering new bugs in regex engines?}
    \end{researchquestions}
\end{researchquestions}

\section{Testing Practices in Regex Engines}\label{sec:testing_practices}
Many programming languages provide built-in regex support through their standard libraries, while others delegate to external open-source implementations.
To characterize the contemporary regex engine landscape, we conducted a systematic engineering review of 22 widely used engines across these two categories.

\myparagraph{Methodology}
  For \textit{first-party engines} (standard library implementations), we surveyed the top 10 programming languages by GitHub push activity~\cite{githut2024}: Python, Java, JavaScript, TypeScript, C, C++, PHP, Go, Ruby, and C\#.
  After consolidating languages that share the same compiler or runtime infrastructure (\ie JavaScript/TypeScript and C/C++), the list reduced to eight distinct engines; we therefore added Shell and Rust, the next highest-ranked languages, to obtain 10 first-party engines.
  For \textit{third-party engines}, we surveyed open-source regex engines on GitHub with at least 500 stars, yielding 11 additional engines: PCRE2, RE2, Oniguruma, Onigmo, Hyperscan, TRE, mrab-regex, regexp2, fancy-regex, sregex, and tiny-regex-c.
  We additionally included PCRE's first version, given its historical influence on regex semantics.

  To compile a representative feature set, we aggregated regex constructs referenced in at least two of four authoritative sources~\cite{davisImpactDefeatRegular2020,regular_expressions_info_flavors,pcre_manual,opengroup_posix_regex_ch9}, ensuring coverage of commonly expected regex features.
  We then verified the presence or absence of each construct in every engine by consulting official documentation and, where documentation was ambiguous or incomplete, inspecting source code directly.

  \Cref{tab:engine-comparison} summarizes our review across 68 syntactic features spanning 13 categories.
  We emphasize that this survey captures feature availability based on syntax support; even when two engines support the same construct, subtle semantic differences may exist~\cite{linguaFranca}.

\subsection{The Testing Challenge}\label{sec:challenge}

Current literature provides no guidance on \textit{``how should we test regex engines?''} or \textit{``how people test regex engines?''}.
To understand the state of practice, we conducted an engineering review of the testing infrastructure in the 22 regex engines from \cref{subsec:regex_engines}.

\myparagraph{Methodology}
For each engine, we inspected the test suite in its source repository and recorded all distinct testing strategies employed.
From the full set of observed techniques, we retained those that appear in at least two engines to capture common practice, yielding 12 techniques.
\Cref{tab:testing-comparison} summarizes our findings.

\vspace{-2mm}
\begin{table}[ht]
\centering
\caption{
Testing techniques employed by 22 regex engines, among those used in $\geq$ 2 engines.
}
\vspace{-2mm}
\scalebox{0.75}{%
\begin{threeparttable}
\label{tab:testing-comparison}
\footnotesize
\small
\renewcommand{\arraystretch}{0.8}
\setlength{\tabcolsep}{7.5pt}
\input{tables/table_2}
\begin{tablenotes}
\footnotesize
\item \cmark=Present, \xmark=Not present or not explicitly documented.
\vspace{-4mm}
\end{tablenotes}
\end{threeparttable}
}%
\end{table}

\vspace{-2mm}
\myparagraph{Observations}
Six techniques---unit testing, regression testing, performance testing, boundary testing, integration testing, and negative testing---are universally adopted across all 22 engines.
While foundational, these techniques are among the standard testing practices for software projects; they demand substantial manual effort to achieve comprehensive coverage in the context of regex features and edge cases.

Differential testing is employed by 11 engines (50\%) as an attempt to establish a semantic oracle.
When an engine's output diverges from a reference implementation, developers investigate whether the discrepancy indicates a bug.
PCRE is the most widely used oracle (used by RE2, Hyperscan, sregex, and regexp2), followed by Perl (used by PCRE, PCRE2 and sregex).

Fuzzing is employed by 10 engines (45\%) to maximize code coverage and discover edge cases that manual testing misses.
However, the sophistication of fuzzing approaches varies considerably.
Only two engines employ grammar-aware fuzzing:
JS/TS's Irregexp and Rust's regex crate.
Most engines rely instead on mutation-based fuzzing (\eg via LibFuzzer~\cite{libfuzzer}, OSS-Fuzz~\cite{ossfuzz_pcre2_2026}) that generates inputs through random byte-level mutations.
Without structural awareness, such mutations are unlikely to produce syntactically valid patterns, limiting coverage of matching logic.

\subsubsection{The Oracle Problem}\label{sec:oracle}

The \textit{test oracle problem}~\cite{barr2015oracle} refers to the difficulty of determining whether a program's output is correct.
For regex engine testing, different testing techniques employ different oracles. Here, we summarize our observations.

\myparagraph{Developer-Defined Oracles}
Unit tests and regression tests rely on manually specified expected outputs.
Developers write assertions such as \textit{``pattern \code{a+} on input \code{aaa} should match \code{aaa}.''}
While precise, this approach scales poorly: exhaustively specifying expected behavior for all combinations of regex features and edge cases is infeasible.

\myparagraph{Crash Oracles}
Fuzzing paired with runtime sanitizers
effectively detects memory safety violations and undefined behavior.
When the engine crashes or a sanitizer reports an error, the oracle is clear: a bug exists.
However, crash oracles cannot detect \textit{semantic} bugs---incorrect match results that do not trigger crashes.

\myparagraph{Differential Oracles}
Differential testing uses a reference implementation as the oracle: if the engine under test produces different output than the reference, a potential bug is flagged.

However, as discussed in \cref{subsec:regex_engines}, regex engines implement different dialects with intentional semantic differences.
Consequently, differential testing generates many \textit{false positives}---discrepancies that reflect expected dialect divergence rather than bugs.

Specifically, Davis~\etal~\cite{davis2019testing} cataloged three classes of such divergence: (1)~same syntax mapped to different features (\eg \verb|\h| matches a tab in PCRE but the literal \texttt{"h"} in Java), (2)~same syntax with different semantics (\eg \verb|^a| on \verb|"x\na"| matches in Ruby but not JavaScript), and (3)~subtle behavioral differences in shared features (\eg \verb|((a*)+)| on \verb|"aa"| captures \verb|""| in Java but \verb|"aa"| in Python).

In all three cases, a true semantic bug would be indistinguishable from expected dialect variation without deep knowledge and manual triage.
This limitation motivates our use of metamorphic testing~(\cref{subsec:oracle_suite}): we need an oracle that validates the semantics within a single engine, independent of cross-implementation comparison.

\begin{rqanswer}[RQ1]
Regex engines rely on \textit{manual testing} (unit, regression, boundary, and negative tests).
\textit{Differential testing} is used by 50\% of engines but suffers from dialect divergence.
\textit{Fuzzing} is employed by 45\% of engines, yet only 2 of 22 engines employ grammar-aware fuzzing for valid inputs.
Overall, current practices lack systematic input generation and a dialect-independent semantic oracle.
\end{rqanswer}

\section{Bugs and Vulnerabilities in Regex Engines}\label{sec:bugs_vulns_in_engines}

We analyze regex engine bugs from their issue trackers (\cref{subsec:bug_review}) and security vulnerabilities from CVE records (\cref{subsec:cves}).

\subsection{Review of Regex Engine Bugs}\label{subsec:bug_review}

To understand the prevalence and nature of bugs in regex engines, we conducted a systematic review of issues from 11 of the third-party regex engine implementations from \cref{subsec:regex_engines}.\footnote{PCRE's issue tracker is no longer available to the public. Hence, we excluded it.}
At the current stage of our research, we focus on third-party engines.
Our future work (\cref{sec:JournalNext}) will also include first-party regex engines and examine failure modes more deeply.

\subsubsection{Methodology}

We collected 2,342 issue reports across the 11 engines via the GitHub API and classified each using an LLM-based analysis pipeline using OpenAI's \texttt{gpt-5.1-mini} model.
Then, we manually verified that each classification output conformed to our predefined schema.
The system prompt we used for classifying regex engine issue reports is available in our artifact (\cref{sec:data-availability}).

For each issue, our classification pipeline determines:
  (1)~whether it represents a \emph{real bug}---a confirmed defect in the engine's implementation, excluding user misunderstandings, feature requests, and API misuse;
  (2)~the \emph{bug type}:
    \ul{semantic} (incorrect match results, whether identified against a specification, a prior version, another engine, or user expectations),
    \ul{crash} (abnormal termination, \eg panics, uncaught exceptions),
    \ul{memory} (leaks, buffer errors),
    \ul{performance} (catastrophic backtracking, timeouts),
    \ul{documentation} (incorrect or misleading docs),
    or
    \ul{other} (confirmed bugs not fitting prior categories, \eg build failures, installation issues);
  and
  (3)~the \emph{testing strategy} used to discover the bug (\eg manual review (user-reported), fuzzing, differential testing).

\subsubsection{Results}

Of the 2,342 issues analyzed, our classification pipeline labeled 1,007 as real bugs (43.0\%), spanning issue reports from 2011 to 2025 (91.8\% of the issue reports are from between 2015 and 2025).
The proportion classified as bugs varies by engine, from 8.7\% in sregex to 59.3\% in Oniguruma. %
Detailed per-engine statistics are available in our artifact (\cref{sec:data-availability}).

\begin{figure}[!ht]
  \centering
  \includegraphics[width=0.70\columnwidth]{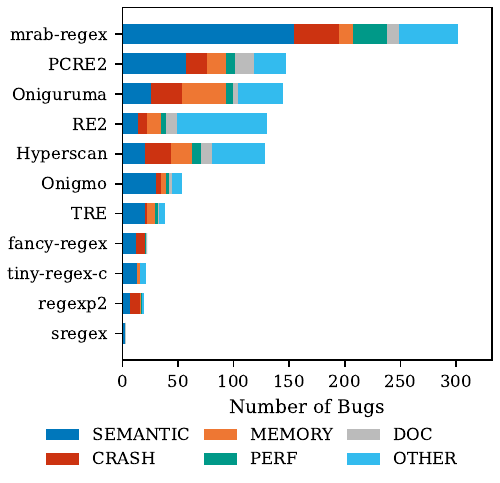}
  \vspace{-4mm}
  \caption{
  Distribution of bug types across 11 third-party regex engines.
  Semantic bugs form a plurality (35.3\%) of all bugs.
  }
  \label{fig:bug-types}
\end{figure}

\Cref{fig:bug-types} shows the distribution of bug types across engines.
The three most common bug categories overall are semantic bugs at 35.3\% (355), followed by crash bugs at 14.3\% (144) and memory-related bugs at 11.4\% (115).
While semantic bugs form the largest single category across all engines combined, the distribution varies by engine; for instance, RE2 and Hyperscan show higher proportions of memory and crash bugs.

\vspace{-3mm}
\vspace{-2mm}

According to our analysis, 82.1\% of bugs were reported by users, while only 6.3\% were discovered via fuzzing and 5.4\% through differential testing.\footnote{Bugs found internally by engine developers may be fixed without public issue reports. However, we assume significant defects---whether found internally or externally---to be documented in issue trackers.}
This discovery gap highlights an opportunity for automated testing to uncover bugs before users encounter them in production.
55.5\% of the identified bugs were subsequently fixed.

\subsection{Review of Regex Engine CVEs}\label{subsec:cves}

To complement our bug study with an analysis of security vulnerabilities, we collected CVEs (Common Vulnerabilities and Exposures) associated with regex engines from the NIST National Vulnerability Database (NVD).

\subsubsection{Methodology}

For each regex engine listed in \cref{subsec:regex_engines} (both first-party and third-party), we queried the NVD API for CVEs that mention the engine, capturing both CVEs assigned directly to the engine and CVEs assigned to downstream projects affected by defects in the engine itself. Next, we manually investigated the query results and manually removed the false positive matches (\eg CVEs from an unrelated software also named RE2).

\subsubsection{Results}

We identified 156 CVEs across 15 of the 22 surveyed engines, spanning the period 2005--2025.
In terms of severity levels, 32 CVEs (21\%) were rated critical and 45 (29\%) were rated high, meaning over half of all regex engine CVEs pose significant security risks.
The remaining vulnerabilities are rated medium severity (19\%) or lack a CVSS score (31\%).

\vspace{-2mm}
\begin{figure}[!ht]
  \centering
  \includegraphics[width=0.70\columnwidth]{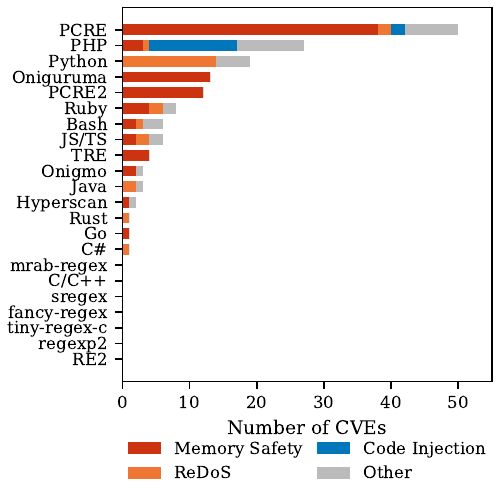}
  \vspace{-4mm}
  \caption{Distribution of CVE types across regex engines.}
  \vspace{-4mm}
  \label{fig:cve-types}
\end{figure}

\Cref{fig:cve-types} shows the distribution of CVE types across engines, categorized by their CWE (Common Weakness Enumeration) classifications.
Memory safety vulnerabilities dominate, accounting for 52\% of all CVEs.
These include buffer overflows (CWE-119), out-of-bounds reads (CWE-125), out-of-bounds writes (CWE-787), integer overflows (CWE-190), and use-after-free errors (CWE-416).
ReDoS-related vulnerabilities (CWE-1333, CWE-400) constitute 17\% of CVEs.
PCRE alone accounts for 50 CVEs (32\% of the total).
PHP's PCRE bindings contribute an additional 27 CVEs, many related to how PHP exposes PCRE functionality.
Seven engines in our survey have zero reported CVEs.%

As motivation, these findings indicate two things.
First, memory safety is the dominant vulnerability class in regex engines, and should be a target of any testing approach.
Second, ``where there is smoke, there is fire''---the abundance of CVEs in the PCRE and PCRE2 engines suggests that systematic testing of other widely-deployed libraries could yield significant security benefits.

\begin{rqanswer}[RQ2]
Semantic bugs are the most common defect type (35\%) overall, followed by crashes (14\%) and memory errors (11\%).
82\% of bugs are user-reported; only 6\% are found via fuzzing.
Among 156 CVEs, memory safety vulnerabilities dominate (52\%), with 50\% rated high or critical severity.
PCRE alone accounts for 32\% of regex engine CVEs.
\end{rqanswer}

\section{Design of \toolname}\label{sec:design}

\begin{figure*}[!ht]
    \centering
    \includegraphics[width=1.5\columnwidth]{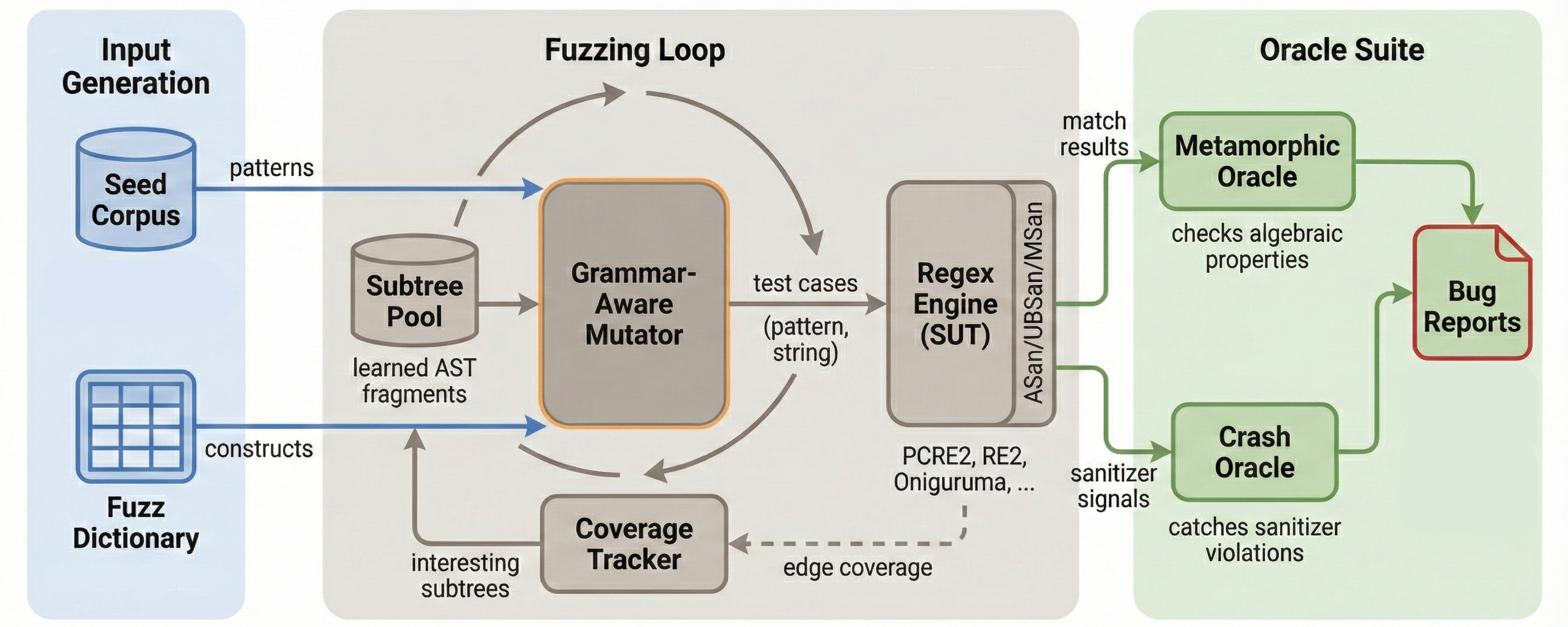}
    \vspace{-2mm}
    \caption{Overview of the \toolname framework.
    The input generation module seeds fuzzing
    and maintains a subtree pool for grammar-aware AST mutations.
    The fuzzing loop generates test cases and tracks edge coverage to guide mutation.
    The oracle suite combines runtime sanitizers for crash detection with metamorphic testing for semantic bug detection. %
    }
    \label{fig:architecture}
\vspace{-4mm}
\end{figure*}

\toolname is a systematic testing framework for regex engines that combines two complementary techniques: (1)~\textit{grammar-aware fuzz\-ing} for generating syntactically valid test inputs that achieve high code coverage, and (2)~\textit{metamorphic testing} for providing a dialect-independent semantic oracle.
\Cref{fig:architecture} illustrates the high-level architecture.

Our utilization of these two techniques is motivated by two findings from our empirical analysis.
\ul{First}, the choice of a fuzzing-based architecture is necessitated by the expansive grammar of modern regex engines (\cref{subsec:regex_engines}); the combinatorial explosion of feature interactions renders exhaustive manual verification cost-prohibitive.
This design choice aligns with the current practices, as $\sim$50\% of the engines surveyed in \cref{sec:testing_practices}
already employ fuzzing to mitigate this scalability challenge. However, these implementations predominantly rely on naive, byte-level mutation strategies; consequently, generated inputs frequently violate syntax rules and are rejected early by the parser, leaving matching logic unexplored.
\ul{Second}, the oracle problem (\cref{sec:oracle}) limits the utility of differential testing:
  dialect divergences may lead to false-positive alerts, necessitating a dialect-independent verification strategy.

\vspace{-8pt}
\subsection{Input Generation}\label{sec:input_generation}

Regex engine testing requires generating two types of inputs: (1)~reg\-ex patterns that exercise diverse engine code paths, and (2)~input strings that trigger matching behavior for those patterns.
Our analysis of existing fuzzing harnesses (\cref{sec:challenge}) shows that most regex engines rely on randomly generated regexes and input strings, which limits coverage of real code paths due to the high likelihood of producing syntactically invalid inputs.
\toolname addresses this through corpus-based seeding with grammar-aware augmentation and pattern-derived string synthesis.

\subsubsection{Regex Pattern Generation}

\toolname can draw seeds from a corpus of real-world regex patterns collected from production software.
We use a dataset of over 500{,}000 unique patterns extracted from open-source projects and internet sources~\cite{cakar2025reuseneedsystematiccomparison}, filtered to retain only patterns that compile successfully on the target engine.
This seeding strategy ensures that initial inputs reflect realistic usage patterns and exercise code paths for production workloads.

\subsubsection{Input String Generation}

Given a regex pattern $r$, \toolname generates input strings to exercise the engine's matching logic.
Our goal is to produce strings that approximate the language $L(r)$ and its complement $\overline{L(r)}$, ensuring coverage of both accepting and rejecting paths through the engine's internal automaton.
For this purpose, we use EGRET~\cite{larsonGeneratingEvilTest2016} to generate a complete set of strings covering all automaton edges, thereby providing coverage guarantees for both matching and non-matching code paths.

\subsection{Oracle Suite}\label{subsec:oracle_suite}

As discussed in~\cref{sec:challenge}, one of the challenges in regex engine testing is the oracle problem: \textit{``given a pattern $r$ and input $s$, how do we determine whether the engine's output is correct?''}
\toolname addresses this through a multi-layered oracle suite combining \textit{metamorphic testing} for semantic correctness and \textit{runtime sanitizers} for safety.

\begin{table}[!ht]
\centering
\caption{Metamorphic relations implemented in \toolname, curated from~\cite{kozenCompletenessTheoremKleene1994,salomaaTwoCompleteAxiom1966}.
All relations apply to K-regexes (Kleene fragments).
$\varepsilon$ denotes the empty string, and $\emptyset$ denotes an always-failing pattern (\eg \code{(?!)}).
``Assertion'' indicates whether the oracle checks match existence (\textit{match}) or also that match spans are identical (\textit{match+span}).
}
\vspace{-2mm}
\label{tab:metamorphic-relations}
\footnotesize
\setlength{\tabcolsep}{8pt}
\input{tables/table_3}
\vspace{-3mm}
\end{table}

\subsubsection{Metamorphic Relations}
While different engines may disagree on specific behaviors, they must all satisfy certain algebraic properties derived from Kleene algebra~\cite{Kleene1951NerveNetsAndRegularLanguages}.
Rather than comparing across engines, we validate that a single engine is consistent with itself under semantics-preserving transformations.
For this, we use metamorphic relations (\cref{subsec:metamorphic_testing}).

\toolname implements 16 metamorphic relations derived from Kleene algebra, organized into five categories.
We compile these relations from Salomaa~\cite{salomaaTwoCompleteAxiom1966} and Kozen~\cite{kozenCompletenessTheoremKleene1994}.
\Cref{tab:metamorphic-relations} presents this catalog.
These relations are \textit{implementation-independent}: any formally correct regex engine, regardless of dialect, must satisfy them.
At the current stage of our research, our MRs cover only the K-regex fragment. Still, \toolname can apply them to K-regex sub-patterns within E-regexes, enabling partial validation of extended patterns.
We are planning to expand our MR catalog for E-regexes too (\cref{sec:JournalNext}).

\myparagraph{Metamorphic Transformation and Validation}

Each metamorphic relation is implemented as an AST visitor that identifies applicable patterns and applies the transformation.
For example, we do not apply Kleene star rules if $r$ does not use the $*$ operator.
\Cref{alg:metamorphic} presents the algorithm.

\noindent
\begin{minipage}{\columnwidth}
\vspace{-1mm}
\begin{algorithm}[H]
\caption{Metamorphic Testing for Regexes}
\label{alg:metamorphic}
\begin{algorithmic}[1]
\small
\REQUIRE Pattern $r$, Input $s$, Engine $E$, Relations $\mathcal{MR}$
\ENSURE Bug report or $\bot$
\FOR{each relation $\textit{mr} \in \mathcal{MR}$}
    \STATE \textbf{if} not $\textit{mr}.\textsc{Precondition}(r)$ \textbf{then continue}
    \STATE $r' \leftarrow \textit{mr}.\textsc{Transform}(r)$
    \STATE $\textit{result}_\textit{base} \leftarrow E.\textsc{Search}(r, s)$;\, $\textit{result}_\textit{variant} \leftarrow E.\textsc{Search}(r', s)$
    \IF{not $\textit{mr}.\textsc{Assert}(\textit{result}_\textit{base}, \textit{result}_\textit{variant})$}
        \RETURN \textsc{Bug}($\textit{mr}$, $r$, $r'$, $s$)
    \ENDIF
\ENDFOR
\RETURN $\bot$
\end{algorithmic}
\end{algorithm}
\vspace{-1mm}
\end{minipage}%

\subsubsection{Sanitizer-Based Oracles}

While metamorphic testing is effective at detecting \textit{semantic} bugs (i.e., incorrect match results), it is not intended to uncover undefined behaviors (\eg crashes) or other violations (\eg memory-safety issues) that do not affect program output.
Together, these issues characterize the current landscape of regex engine bugs and vulnerabilities, alongside semantic bugs, as discussed in \cref{sec:bugs_vulns_in_engines}.

To detect safety violations, \toolname integrates three standard runtime sanitizers:
  AddressSanitizer (ASan)~\cite{Serebryany2012AddressSanitizer}
  and
   MemorySanitizer (MSan)~\cite{Stepanov2015MemorySanitizer}
  and
  UndefinedBehaviorSanitizer (UBSan)~\cite{ubsan} for undefined behaviors such as signed integer overflows and null pointer dereferences, %
When a sanitizer detects a violation, this comprises a crash oracle, accompanied by diagnostic information.

\subsection{Guidance \& Mutations}\label{sec:guidance}
While corpus seeding and dictionary injection provide diverse initial inputs, systematic exploration of engine internals requires \textit{coverage-guided mutation}---evolving patterns based on feedback about which code paths have been exercised.
\toolname combines coverage tracking with \textit{context-aware} grammar-based mutation to iteratively discover inputs that reach deep engine states.

\subsubsection{Coverage Tracking}

\toolname tracks edge coverage in regex engine binaries.
The coverage tracker maintains a bitmap of observed edges, updated after each engine execution.
When a test input triggers previously-unseen edges, it is marked as ``interesting'' and used to guide subsequent mutation.

\subsubsection{AST-Based Subtree Replacement}

For deep exploration of engine internals, \toolname performs grammar-aware AST-based mutation on patterns that trigger new coverage.
The mutator parses patterns into abstract syntax trees and performs structured modifications---replacing subtrees with semantically compatible alternatives from a learned pool.

Influenced from~\cite{8811923}, each subtree in the pool is indexed by its \textit{syntactic context}---the types of its ancestor nodes in the original AST.
When selecting a replacement for a mutation target, \toolname retrieves subtrees that appeared in similar contexts, ensuring semantic compatibility (\eg a quantifier body is replaced with another valid atom, not an anchor).
\Cref{alg:mutation} presents the mutation algorithm.
An optimization we applied is \textit{copy-on-write} AST manipulation: rather than deep-copying the entire tree for each mutation, we copy only the path from root to the mutation target, leaving untouched subtrees as shared references.

\noindent
\begin{minipage}{\columnwidth}
\vspace{-2mm}
\begin{algorithm}[H]
\caption{Grammar-Aware Mutation}
\label{alg:mutation}
\begin{algorithmic}[1]
\small
\REQUIRE Pattern $p$, Subtree Pool $\mathcal{P}$
\ENSURE Mutated pattern $p'$
\STATE $\textit{ast} \leftarrow \textsc{Parse}(p)$
\STATE $\textit{nodes} \leftarrow \textsc{CollectInternalNodes}(\textit{ast})$
\STATE $\textit{target} \leftarrow \textsc{RandomSelect}(\textit{nodes})$
\STATE $\textit{ctx} \leftarrow \textsc{ExtractContext}(\textit{target})$
\STATE $\textit{repl} \leftarrow \mathcal{P}.\textsc{Get}(\textsc{Type}(\textit{target}), \textit{ctx})$
\STATE $\textit{ast}' \leftarrow \textsc{Replace}(\textit{ast}, \textit{target}, \textit{repl})$
\RETURN \textsc{Serialize}($\textit{ast}'$)
\end{algorithmic}
\end{algorithm}
\vspace{-5mm}
\end{minipage}

\subsubsection{Coverage-Driven Subtree Pool}
The subtree pool stores AST fragments organized by node type and syntactic context, enabling efficient lookup of compatible replacements.
Rather than populating the pool statically, \toolname updates it \textit{dynamically} based on coverage feedback: when a mutated pattern triggers new code paths, its subtrees are extracted and added to the pool.
This coverage-driven approach ensures the pool accumulates ``interesting'' subtrees---those that exercise diverse engine behaviors.

\section{Preliminary Evaluation of \toolname}\label{sec:preliminary_evaluation}

At the current prototyping stage of \toolname, we conducted a preliminary evaluation to explore its capabilities, particularly for RQ3b (coverage and fault detection). We compare \toolname against two baselines representing current practice in regex engine fuzzing.

\myparagraph{Baselines}
We compare three fuzzing strategies:

\begin{enumerate}
    \item \textit{B1: Naive Fuzzing}. We use the OSS-Fuzz~\cite{ossfuzz_pcre2_2026} harness for PCRE with LibFuzzer's default mutation strategy, which applies random bit-flips, byte insertions, and deletions without awareness of regex syntax.
    \item \textit{B2: Grammar-Aware Fuzzing}. We adapt the grammar-based fuzzing approach from JS/TS (V8)'s Irregexp test suite, which generates patterns by random production rule expansion and replaces AST subtrees with subtrees of the same node type. While this ensures syntactic validity, type-only matching ignores syntactic context and can produce semantically malformed patterns (\eg replacing a quantifier body with an anchor). This represents the state of practice for structure-aware fuzzing (\cref{sec:testing_practices}).
    \item \textit{\toolname}. Our approach bootstraps fuzzing with 1,000 seed patterns randomly sampled from our 500K regex corpus and applies coverage-guided, context-aware AST mutation that indexes subtrees by their syntactic context (\cref{sec:guidance}).
\end{enumerate}

\myparagraph{Setup}

We evaluate all three strategies on PCRE v8.45, compiled with ASan and UBSan.
Each configuration runs for 1 hour on a dedicated allocation of 4 cores from an Intel Xeon W-2295 CPU with 188GB RAM.
We measure unique edges covered and the number of unique bugs discovered.

\begin{figure}[!ht]
\centering
\includegraphics[width=0.70\columnwidth]{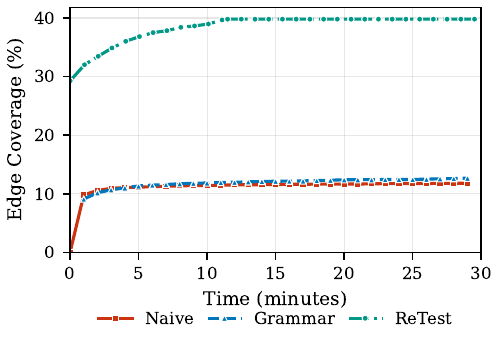}
\vspace{-4mm}
\caption{Edge coverage over time for three strategies on PCRE v8.45.}
\label{fig:coverage-comparison}
\end{figure}

\myparagraph{Results}
\Cref{fig:coverage-comparison} shows edge coverage over time.

\vspace{-3mm}

\toolname achieves 40.12\% edge coverage, compared to 12.66\% for V8-style grammar-aware fuzzing and 11.81\% for OSS-Fuzz-style naive fuzzing. ReTest's corpus seeds provide an obvious advantage of 29.4\% near-immediate initial coverage, while other baselines start near zero.
Notably, \toolname's context-aware mutations contribute an additional $\sim$11 percentage points (from 29\% to 40\%), which is comparable to the total coverage achieved by baselines---despite diminishing returns at higher coverage levels where novel edges become increasingly sparse.

The iteration counts reflect different efficiency characteristics.
Naive fuzzing executes $\sim$156M iterations; grammar-aware fuzzing reduces this to $\sim$5.3M; \toolname executes $\sim$40K iterations.
\toolname's coverage plateaus around minute 12, indicating room for improvement for the full submission of our work (\cref{sec:JournalNext}).

\myparagraph{Bugs Found}

During our prototyping of \toolname, we discovered three bugs in PCRE v8.45.
Two patterns trigger heap memory corruption: one involving the \texttt{\textbackslash{}C} escape sequence in UTF-8 mode (\texttt{\textbackslash{}P\{Z\}+:gra]sh*:p(\textbackslash{}CR)(?s)}), and another combining \texttt{\textbackslash{}C} with property-like syntax (\texttt{\textbackslash{}p\{Han\}*\textbackslash{}n*\textbackslash{}t*\textbackslash{}.\textbackslash{}CP\{\}*\textbackslash{}n*\textbackslash{}t*body}) (CWE-122, CWE-787).
Both cause heap corruption at compile-time or JIT study-time.
A third pattern involving zero-quantified recursion \texttt{((x\{1,3\}|\textbackslash{}p\{L\}++|(([\textasciicircum{}>]*(?1)\{0\}(?1)?)))+)}  triggers a global buffer overflow during compilation (CWE-125).
These bugs are related to some known vulnerabilities but not the same (\eg CVE-2015-2325, CVE-2015-2328).
Neither of the other two baselines revealed any bugs in our experiments.
Detailed analysis is available in our artifact (\cref{sec:data-availability}).

\vspace{-2mm}
\section{Future Work}
\label{sec:JournalNext}

We regard RQ1 (\cref{sec:testing_practices}) and RQ2 (\cref{sec:bugs_vulns_in_engines}) as close to mature contributions. We plan to expand our bug analysis to include \emph{first-party} regex engines, which requires filtering to isolate regex engine-specific defects from broader runtime issues in repositories like CPython and OpenJDK.
We will also improve the soundness of our method through sampling for deeper manual analysis.
Additionally, we will document the code coverage achieved by each testing strategy observed in~\cref{tab:testing-comparison} (\eg unit tests, regression tests) to quantify the effectiveness of current practices and identify coverage gaps that \toolname can address.

For RQ3~(\cref{sec:design,sec:preliminary_evaluation}), future work remains. Our metamorphic relation catalog (\cref{tab:metamorphic-relations}) currently covers only K-regexes. Drawing on recent formalization efforts~\cite{berglundSemanticsAtomicSubgroups2017,berglundRegularExpressionsLookahead2021,chattopadhyayVerifiedEfficientMatching2025,fujinamiEfficientMatchingMemoization2024}, we will extend the catalog to E-regex constructs, enabling validation of a broader class of patterns.
Additionally, our preliminary evaluation of \toolname reveals coverage plateaus at $\sim$40\% edge coverage mark; we will investigate the reasons in-depth and develop strategies to break through these limitations.
Our current conjecture is that automaton coverage is an inadequate heuristic for selecting strings to accompany the regexes.

The full evaluation will address all RQ3 sub-questions comprehensively. For RQ3a, we will compare our metamorphic oracle against differential testing, measuring false positive rates.
For RQ3b, we will scale fuzzing experiments to all engines with 24-hour campaigns, comparing \toolname against naive and grammar-aware baselines with statistical rigor.
We will also measure \toolname's and other baselines' ability to replicate known bugs and vulnerabilities from our review (\cref{sec:bugs_vulns_in_engines}).
For RQ3c, we will report all discovered bugs with responsible disclosure and maintainer response tracking.

Finally, we will ablate the contributions of \toolname parameters, \eg the initial corpus size and subtree pool size. %

\section{Conclusion}

Regular expression engines are critical infrastructure, yet our empirical analysis reveals significant testing gaps: 82\% of bugs are user-reported rather than caught by systematic testing, and existing approaches lack a dialect-independent semantic oracle.
We presented \toolname, a framework that addresses these gaps by combining grammar-aware fuzzing with metamorphic testing grounded in Kleene algebra.
Our preliminary evaluation on PCRE v8.45 demonstrates that \toolname achieves 3$\times$ higher edge coverage than in-practice approaches and has already uncovered three memory safety bugs.
We plan to extend this work with a comprehensive evaluation across 22 regex engines, an expanded metamorphic relation catalog for E-regex features, and ablation studies to guide future regex engine testing research.

\section*{Data Availability}\label{sec:data-availability}
\phantomsection
\addcontentsline{toc}{section}{Data Availability}
An artifact associated with this paper is available at \url{https://github.com/PurdueDualityLab/ReTest}.

\ifANONYMOUS
\else
\begin{acks}
This work was supported by the \grantsponsor{YRXVL4JYCEF5}{US National Science Foundation (NSF)}{} under grants
SaTC-2135156
and
SaTC-2135157.
We thank August Shi for his valuable feedback at an early stage of this work.
\end{acks}
\fi

\printbibliography

\end{document}

%% file: typesetting.tex
\usepackage{packages/acmart-taps}
\usepackage[export]{adjustbox} %
\usepackage{amsmath}
\usepackage{amsfonts}
\usepackage{array}
\usepackage[USenglish]{babel}
\usepackage{packages/bootstrapicons}
\usepackage{colortbl}
\usepackage{cleveref}
\usepackage{enumerate}
\usepackage{float}
\usepackage{forest}
\usepackage[htt]{hyphenat}
\usepackage{makecell}
\usepackage{multirow}
\usepackage{textgreek}
\usepackage{soul}
\usepackage{pifont}
\usepackage{subcaption} %
\usepackage{url}
\usepackage[table]{xcolor}
\usepackage{xspace}
\usepackage{wrapfig}
\usepackage{algorithm}
\usepackage{algorithmic}
\usepackage{paracol}
\usepackage{tabularx}
\usepackage{threeparttable} %
\usepackage{packages/picins}

\usepackage{tikz} %
\usetikzlibrary{automata} %
\usetikzlibrary{positioning} %
\usetikzlibrary{arrows} %
\usetikzlibrary{calc} %
\usetikzlibrary{patterns} %
\usetikzlibrary{snakes} %
\usetikzlibrary{shapes} %
\usetikzlibrary{arrows.meta} %

\newcommand{\code}[1]{\texttt{{\small #1}}}

\newcommand{\ie}{\textit{i.e.,\ }}
\newcommand{\eg}{\textit{e.g.,\ }}
\newcommand{\etal}{\textit{et al.\ }}
\newcommand{\etals}{\textit{et al.}'s\ }
\newcommand{\adhoc}{\textit{ad hoc}\xspace}

\newcommand{\vs}{{\em vs.}\xspace}

\usepackage{mdframed}
\mdfsetup{skipabove=0.5\topskip,skipbelow=0.5\topskip,align=center}

\usepackage{amsthm}

\usepackage{enumitem}
\setlist[itemize]{leftmargin=*,noitemsep,topsep=0pt}
\setlist[enumerate]{leftmargin=*}

\newcommand{\cmark}{\textcolor{green!60!black}{\ding{51}}}
\newcommand{\xmark}{\textcolor{red!70!black}{\ding{55}}}
\newcommand{\qmark}{\fontfamily{cyklop}\selectfont\tiny\textit{?}}

\newlist{researchquestions}{enumerate}{2}
\setlist[researchquestions,1]{
  label=\textbf{RQ\arabic*},
  leftmargin=*,
  ref=RQ\arabic*
}

\setlist[researchquestions,2]{
  label=\textbf{RQ\arabic{researchquestionsi}\alph*},
  ref=RQ\arabic{researchquestionsi}\alph*,
  leftmargin=*
}

\crefformat{section}{\S#2#1#3}
\crefmultiformat{section}{\S#2#1#3}{, \S#2#1#3}{, and \S#2#1#3}{}
\crefname{figure}{Figure}{Figures}
\crefname{appendix}{Appendix}{Appendices}
\crefname{table}{Table}{Tables}
\crefname{algorithm}{Algorithm}{Algorithms}
\crefname{listing}{Listing}{Listings}
\crefname{lemma}{Lemma}{Lemmata}
\crefname{equation}{Eqt.}{Eqts.}
\crefformat{Grammar}{Grammar #1}

\newcommand{\myparagraph}[1]{\paragraph{#1}}

\ifDEBUG
  \PassOptionsToPackage{draft}{commenting}
\else
  \PassOptionsToPackage{final}{commenting}
\fi

\usepackage[nompar]{packages/commenting}

\declareauthor{EB}{Ethan}{green}
\declareauthor{CS}{Charlie}{magenta}
\declareauthor{JD}{Jamie}{blue}
\declareauthor{SC}{Sophie}{pink}
\declareauthor{BC}{Berk}{orange}
\declareauthor{DY}{Dongyoon}{cyan}

\authorcommand{EB}{comment}
\authorcommand{CS}{comment}
\authorcommand{JD}{comment}
\authorcommand{SC}{comment}
\authorcommand{BC}{comment}
\authorcommand{DY}{comment}

\onlyauthors{BC,CS,JD,EB,SC,DY} %

\newcommand{\toolname}{{\textsc{ReTest}}\xspace}

\usepackage{xpatch}
\xpatchbibmacro{date}
  {\printtext[parens]{\printdate}}
  {\iffieldundef{date}{}{%
     \printtext[parens]{\printdate}}}
  {}{\typeout{Failed to patch date macro.}}

\ifARXIV
\makeatletter

\xpatchcmd{\@mkbibcitation}
  {ACM, New York, NY, USA}
  {}
  {}
  {\PackageWarning{acmart-patch}{Patching \@mkbibcitation failed!}}

\xpatchcmd{\@mkbibcitation}
  {\@article@string\unskip, \ref{TotPages}}
  {\@article@string\unskip\ref{TotPages}}
  {}
  {\PackageWarning{acmart-patch}{Comma patch failed!}}
  
\makeatother
\else
\fi

\usepackage{packages/coloredtheorem}
\tcbuselibrary{breakable}

\newtcbtheorem{definition}{Definition}{
  breakable,
  enhanced,
  colframe=blue!50!black,
  colback=white,
  colbacktitle=blue!50!black,
  coltitle=white,
  fonttitle=\bfseries,
  boxrule=0.5pt,
  arc=2pt,
  left=4pt, right=4pt, top=4pt, bottom=4pt
}{def}

\newtcbtheorem{lemma}{Lemma}{
  breakable,
  enhanced,
  colframe=yellow!60!black,
  colback=white,
  colbacktitle=yellow!60!black,
  coltitle=white,
  fonttitle=\bfseries,
  boxrule=0.5pt,
  arc=2pt,
  left=4pt, right=4pt, top=4pt, bottom=4pt
}{lem}

\newtcbtheorem{example}{Example}{
  breakable,
  enhanced,
  colframe=gray!60,
  colback=white,
  colbacktitle=gray!60,
  coltitle=white,
  fonttitle=\bfseries,
  boxrule=0.5pt,
  arc=2pt,
  left=4pt, right=4pt, top=4pt, bottom=4pt
}{ex}

\newtcbtheorem{exploit}{Exploit}{
  breakable,
  enhanced,
  colframe=gray!60,
  colback=white,
  colbacktitle=gray!60,
  coltitle=white,
  fonttitle=\bfseries,
  boxrule=0.5pt,
  arc=2pt,
  left=4pt, right=4pt, top=4pt, bottom=4pt
}{exp}

\newtcbtheorem{pattern}{Pattern}{
  breakable,
  enhanced,
  colframe=purple!50!black,
  colback=white,
  colbacktitle=purple!50!black,
  coltitle=white,
  fonttitle=\bfseries,
  boxrule=0.5pt,
  arc=2pt,
  left=4pt, right=4pt, top=4pt, bottom=4pt
}{pat}

\newtcolorbox{rqanswer}[1][]{
  breakable,
  enhanced,
  colframe=teal!70!black,
  colback=white,
  colbacktitle=teal!70!black,
  coltitle=white,
  fonttitle=\bfseries,
  title={Our Preliminary Answer to #1},
  boxrule=0.5pt,
  arc=2pt,
  left=4pt, right=4pt, top=4pt, bottom=4pt
}

\crefname{tcb@cnt@definition}{Definition}{Definitions}
\crefname{tcb@cnt@lemma}{Lemma}{Lemmata}
\crefname{tcb@cnt@example}{Example}{Examples}
\crefname{tcb@cnt@pattern}{Pattern}{Patterns}
\Crefname{tcb@cnt@definition}{Definition}{Definitions}
\Crefname{tcb@cnt@lemma}{Lemma}{Lemmata}
\Crefname{tcb@cnt@example}{Example}{Examples}
\Crefname{tcb@cnt@pattern}{Pattern}{Patterns}
\Crefname{tcb@cnt@exploit}{Exploit}{Exploits}

%% file: tables/table_1.tex
\begin{tabular}{@{}ll|c|ccc|ccccccccc|cccccccccc|cc|ccccc|c|ccccccc|cccccccc|ccccc|ccccc|cccc|ccc|cccc@{}}
\toprule
& & & \multicolumn{3}{c|}{\textbf{Literals}} & \multicolumn{9}{c|}{\textbf{Characters}} & \multicolumn{10}{c|}{\textbf{Character Types}} & \multicolumn{2}{c|}{\textbf{Unicode}} & \multicolumn{5}{c|}{\textbf{Character Classes}} & \textbf{Alt.} & \multicolumn{7}{c|}{\textbf{Quantifiers}} & \multicolumn{8}{c|}{\textbf{Anchors}} & \multicolumn{5}{c|}{\textbf{Capturing}} & \multicolumn{5}{c|}{\textbf{Options}} & \multicolumn{4}{c|}{\textbf{L'arounds}} & \multicolumn{3}{c|}{\textbf{Backrefs.}} & \multicolumn{4}{c}{\textbf{Advnc'd}} \\
& \textbf{Engine} & \rotatebox{90}{\textbf{Nature of the Engine}}
& \rotatebox{90}{\texttt{x}} & \rotatebox{90}{\texttt{\textbackslash x}} & \rotatebox{90}{\texttt{\textbackslash Q...\textbackslash E}}
& \rotatebox{90}{\texttt{\textbackslash a}} & \rotatebox{90}{\texttt{\textbackslash cx}} & \rotatebox{90}{\texttt{\textbackslash e}} & \rotatebox{90}{\texttt{\textbackslash f}} & \rotatebox{90}{\texttt{\textbackslash n}} & \rotatebox{90}{\texttt{\textbackslash r}} & \rotatebox{90}{\texttt{\textbackslash t}} & \rotatebox{90}{\texttt{\textbackslash 0dd}--\texttt{\textbackslash o\{ddd...\}}} & \rotatebox{90}{\texttt{\textbackslash xhh}--\texttt{\textbackslash x\{hhh...\}}}
& \rotatebox{90}{\texttt{.}} & \rotatebox{90}{\texttt{\textbackslash C}} & \rotatebox{90}{\texttt{\textbackslash d --- \textbackslash D}} & \rotatebox{90}{\texttt{\textbackslash h --- \textbackslash H}} & \rotatebox{90}{\texttt{\textbackslash s --- \textbackslash S}} & \rotatebox{90}{\texttt{\textbackslash v --- \textbackslash V}} & \rotatebox{90}{\texttt{\textbackslash w --- \textbackslash W}} & \rotatebox{90}{\texttt{\textbackslash N}} & \rotatebox{90}{\texttt{\textbackslash R}} & \rotatebox{90}{\texttt{\textbackslash X}}
& \rotatebox{90}{\texttt{\textbackslash p\{xx\}}} & \rotatebox{90}{\texttt{\textbackslash P\{xx\}}}
& \rotatebox{90}{\texttt{[...]}} & \rotatebox{90}{\texttt{[$\textasciicircum${}...]}} & \rotatebox{90}{\texttt{[x-y]}} & \rotatebox{90}{\texttt{[[:xxx:]]}} & \rotatebox{90}{\texttt{[[:$\textasciicircum${xxx}:]]}}
& \rotatebox{90}{\texttt{expr|expr|expr...}}
& \rotatebox{90}{\texttt{? --- * --- +}} & \rotatebox{90}{\texttt{?? --- *? --- +?}} & \rotatebox{90}{\texttt{?+ --- *+ --- ++}} & \rotatebox{90}{\texttt{\{n\}}} & \rotatebox{90}{\texttt{\{n,m\}}} & \rotatebox{90}{\texttt{\{n,\} --- \{,m\}}} & \rotatebox{90}{\texttt{(?>...)}}
& \rotatebox{90}{\texttt{\^{}}} & \rotatebox{90}{\texttt{\$}} & \rotatebox{90}{\texttt{\textbackslash A}} & \rotatebox{90}{\texttt{\textbackslash Z}} & \rotatebox{90}{\texttt{\textbackslash z}} & \rotatebox{90}{\texttt{\textbackslash G}}
& \rotatebox{90}{\texttt{\textbackslash b --- \textbackslash B}} & \rotatebox{90}{\texttt{\textbackslash K}}
& \rotatebox{90}{\texttt{(...)}} & \rotatebox{90}{\texttt{(?:...)}} & \rotatebox{90}{\texttt{(?|...)}} & \rotatebox{90}{Named capt. groups} & \rotatebox{90}{\texttt{(?\#...)}}
& \rotatebox{90}{\texttt{(?i)}} & \rotatebox{90}{\texttt{(?J)}} & \rotatebox{90}{\texttt{(?m)}} & \rotatebox{90}{\texttt{(?s)}} & \rotatebox{90}{\texttt{(?x)}}
& \rotatebox{90}{\texttt{(?=)}} & \rotatebox{90}{\texttt{(?!)}} & \rotatebox{90}{\texttt{(?<=)}} & \rotatebox{90}{\texttt{(?<!)}}
& \rotatebox{90}{\texttt{\textbackslash n --- \textbackslash gn --- \textbackslash g\{n\}}} & \rotatebox{90}{\texttt{\textbackslash g\textpm n --- \textbackslash g\{\textpm n\}}} & \rotatebox{90}{Named backrefs.}
& \rotatebox{90}{Subroutines \& recursion} & \rotatebox{90}{Conditionals} & \rotatebox{90}{BT Control Verbs} & \rotatebox{90}{Callouts} \\
\midrule
\multicolumn{69}{l}{\textit{First-Party (Language Standard Library)}} \\
\midrule
Python & \texttt{re} & BT
& \cmark & \cmark & \xmark
& \cmark & \cmark & \xmark & \cmark & \cmark & \cmark & \cmark & \cmark & \cmark
& \cmark & \xmark & \cmark & \xmark & \cmark & \xmark & \cmark & \xmark & \xmark & \xmark
& \xmark & \xmark
& \cmark & \cmark & \cmark & \xmark & \xmark
& \cmark
& \cmark & \cmark & \cmark & \cmark & \cmark & \cmark & \cmark
& \cmark & \cmark & \cmark & \cmark & \cmark & \xmark
& \cmark & \xmark
& \cmark & \cmark & \xmark & \cmark & \cmark
& \cmark & \xmark & \cmark & \cmark & \cmark
& \cmark & \cmark & \qmark & \qmark
& \cmark & \xmark & \cmark
& \xmark & \cmark & \xmark & \xmark \\
\rowcolor{gray!10} Java & \texttt{j.u.regex} & BT
& \cmark & \cmark & \cmark
& \cmark & \cmark & \cmark & \cmark & \cmark & \cmark & \cmark & \cmark & \cmark
& \cmark & \xmark & \cmark & \cmark & \cmark & \cmark & \cmark & \xmark & \cmark & \xmark
& \cmark & \cmark
& \cmark & \cmark & \cmark & \xmark$^\ddagger$ & \xmark$^\ddagger$
& \cmark
& \cmark & \cmark & \cmark & \cmark & \cmark & \cmark & \cmark
& \cmark & \cmark & \cmark & \cmark & \cmark & \cmark
& \cmark & \xmark
& \cmark & \cmark & \xmark & \cmark & \cmark
& \cmark & \xmark & \cmark & \cmark & \cmark
& \cmark & \cmark & \cmark & \cmark
& \cmark & \xmark & \cmark
& \xmark & \xmark & \xmark & \xmark \\
JS/TS & \texttt{Irregexp} & BT
& \cmark & \cmark & \xmark
& \xmark & \xmark & \xmark & \cmark & \cmark & \cmark & \cmark & \cmark & \cmark
& \cmark & \xmark & \cmark & \xmark & \cmark & \xmark & \cmark & \xmark & \xmark & \xmark
& \cmark$^\dagger$ & \cmark$^\dagger$
& \cmark & \cmark & \cmark & \xmark & \xmark
& \cmark
& \cmark & \cmark & \xmark & \cmark & \cmark & \cmark & \xmark
& \cmark & \cmark & \xmark & \xmark & \xmark & \qmark$^\ddagger$
& \cmark & \xmark
& \cmark & \cmark & \xmark & \cmark & \xmark
& \cmark & \xmark & \cmark & \cmark & \xmark
& \cmark & \cmark & \cmark & \cmark
& \cmark & \xmark & \cmark
& \xmark & \xmark & \xmark & \xmark \\
\rowcolor{gray!10} C++ & \texttt{<regex>} & BT
& \cmark & \cmark & \xmark
& \xmark & \cmark & \xmark & \cmark & \cmark & \cmark & \cmark & \cmark & \cmark
& \cmark & \xmark & \cmark & \xmark & \cmark & \xmark & \cmark & \xmark & \xmark & \xmark
& \xmark & \xmark
& \cmark & \cmark & \cmark & \cmark & \cmark
& \cmark
& \cmark & \cmark & \xmark & \cmark & \cmark & \cmark & \xmark
& \cmark & \cmark & \xmark & \xmark & \xmark & \xmark
& \cmark & \xmark
& \cmark & \cmark & \xmark & \xmark & \xmark
& \cmark$^\ddagger$ & \xmark & \cmark$^\ddagger$ & \cmark$^\ddagger$ & \cmark$^\ddagger$
& \cmark & \cmark & \xmark & \xmark
& \cmark & \xmark & \xmark
& \xmark & \xmark & \xmark & \xmark \\
Go$^a$ & \texttt{regexp} & Auto
& \cmark & \cmark & \cmark
& \cmark & \cmark & \xmark & \cmark & \cmark & \cmark & \cmark & \cmark & \cmark
& \cmark & \xmark & \cmark & \xmark & \cmark & \xmark & \cmark & \xmark & \xmark & \xmark
& \cmark & \cmark
& \cmark & \cmark & \cmark & \cmark & \cmark
& \cmark
& \cmark & \cmark & -- & \cmark & \cmark & \cmark & --
& \cmark & \cmark & \cmark & \xmark & \cmark & \xmark
& \cmark & \xmark
& \cmark & \cmark & \xmark & \cmark & \xmark
& \cmark & \xmark & \cmark & \cmark & \cmark
& \xmark & \xmark & \xmark & \xmark
& \xmark & \xmark & \xmark
& \xmark & \xmark & -- & \xmark \\
\rowcolor{gray!10} C\# & \texttt{S.T.Regex} & BT
& \cmark & \cmark & \xmark
& \cmark & \cmark & \cmark & \cmark & \cmark & \cmark & \cmark & \cmark & \cmark
& \cmark & \xmark & \cmark & \xmark & \cmark & \xmark & \cmark & \xmark & \xmark & \xmark
& \cmark & \cmark
& \cmark & \cmark & \cmark & \xmark & \xmark
& \cmark
& \cmark & \cmark & \xmark & \cmark & \cmark & \cmark & \cmark
& \cmark & \cmark & \cmark & \cmark & \cmark & \cmark
& \cmark & \xmark
& \cmark & \cmark & \xmark & \cmark & \cmark
& \cmark & \xmark & \cmark & \cmark & \cmark
& \cmark & \cmark & \cmark & \cmark
& \cmark & \xmark & \cmark
& \xmark & \cmark & \xmark & \xmark \\
Rust$^a$ & \texttt{regex} & Auto
& \cmark & \cmark & \xmark
& \cmark & \cmark & \xmark & \cmark & \cmark & \cmark & \cmark & \cmark & \cmark
& \cmark & \xmark & \cmark & \xmark & \cmark & \xmark & \cmark & \xmark & \xmark & \xmark
& \cmark & \cmark
& \cmark & \cmark & \cmark & \cmark & \cmark
& \cmark
& \cmark & \cmark & -- & \cmark & \cmark & \cmark & --
& \cmark & \cmark & \cmark & \xmark & \cmark & \xmark
& \cmark & \xmark
& \cmark & \cmark & \xmark & \cmark & \xmark
& \cmark & \xmark & \cmark & \cmark & \cmark
& \xmark & \xmark & \xmark & \xmark
& \xmark & \xmark & \xmark
& \xmark & \xmark & -- & \xmark \\
\rowcolor{gray!10} PHP$^b$ & \texttt{preg\_*} & BT
& \cmark & \cmark & \cmark
& \cmark & \cmark & \cmark & \cmark & \cmark & \cmark & \cmark & \cmark & \cmark
& \cmark & \cmark & \cmark & \cmark & \cmark & \cmark & \cmark & \cmark & \cmark & \cmark
& \cmark & \cmark
& \cmark & \cmark & \cmark & \cmark & \cmark
& \cmark
& \cmark & \cmark & \cmark & \cmark & \cmark & \cmark & \cmark
& \cmark & \cmark & \cmark & \cmark & \cmark & \cmark
& \cmark & \cmark
& \cmark & \cmark & \cmark & \cmark & \cmark
& \cmark & \cmark & \cmark & \cmark & \cmark
& \cmark & \cmark & \cmark & \cmark
& \cmark & \cmark & \cmark
& \cmark & \cmark & \cmark & \cmark \\
Ruby$^c$ & \texttt{Regexp} & BT
& \cmark & \cmark & \xmark
& \cmark & \cmark & \cmark & \cmark & \cmark & \cmark & \cmark & \cmark & \cmark
& \cmark & \xmark & \cmark & \xmark & \cmark & \xmark & \cmark & \xmark & \cmark & \cmark
& \cmark & \cmark
& \cmark & \cmark & \cmark & \cmark & \cmark
& \cmark
& \cmark & \cmark & \cmark & \cmark & \cmark & \cmark & \cmark
& \cmark & \cmark & \cmark & \cmark & \cmark & \cmark
& \cmark & \cmark
& \cmark & \cmark & \xmark & \cmark & \cmark
& \cmark & \xmark & \cmark & \cmark & \cmark
& \cmark & \cmark & \cmark & \cmark
& \cmark & \cmark & \cmark
& \cmark & \cmark & \xmark & \xmark \\
\rowcolor{gray!10} Shell & \texttt{[[ =}$\sim$\texttt{ ]]} & Auto
& \cmark & \cmark & \xmark
& \xmark & \xmark & \xmark & \xmark & \cmark & \cmark & \cmark & \xmark & \xmark
& \cmark & \xmark & \cmark & \xmark & \cmark & \xmark & \cmark & \xmark & \xmark & \xmark
& \xmark & \xmark
& \cmark & \cmark & \cmark & \cmark & \cmark
& \cmark
& \cmark & \xmark & -- & \cmark & \cmark & \cmark & --
& \cmark & \cmark & \xmark & \xmark & \xmark & \xmark
& \xmark & \xmark
& \cmark & \xmark & \xmark & \xmark & \xmark
& \xmark & \xmark & \xmark & \xmark & \xmark
& \xmark & \xmark & \xmark & \xmark
& \cmark & \xmark & \xmark
& \xmark & \xmark & -- & \xmark \\
\midrule
\multicolumn{69}{l}{\textit{Third-Party Libraries}} \\
\midrule
& PCRE & Hyb
& \cmark & \cmark & \cmark
& \cmark & \cmark & \cmark & \cmark & \cmark & \cmark & \cmark & \cmark & \cmark
& \cmark & \cmark & \cmark & \cmark & \cmark & \cmark & \cmark & \cmark & \cmark & \cmark
& \cmark & \cmark
& \cmark & \cmark & \cmark & \cmark & \cmark
& \cmark
& \cmark & \cmark & \cmark & \cmark & \cmark & \cmark & \cmark
& \cmark & \cmark & \cmark & \cmark & \cmark & \cmark
& \cmark & \cmark
& \cmark & \cmark & \cmark & \cmark & \cmark
& \cmark & \cmark & \cmark & \cmark & \cmark
& \cmark & \cmark & \cmark & \cmark
& \cmark & \cmark & \cmark
& \cmark & \cmark & \cmark & \cmark \\
\rowcolor{gray!10} & PCRE2 & Hyb
& \cmark & \cmark & \cmark
& \cmark & \cmark & \cmark & \cmark & \cmark & \cmark & \cmark & \cmark & \cmark
& \cmark & \cmark & \cmark & \cmark & \cmark & \cmark & \cmark & \cmark & \cmark & \cmark
& \cmark & \cmark
& \cmark & \cmark & \cmark & \cmark & \cmark
& \cmark
& \cmark & \cmark & \cmark & \cmark & \cmark & \cmark & \cmark
& \cmark & \cmark & \cmark & \cmark & \cmark & \cmark
& \cmark & \cmark
& \cmark & \cmark & \cmark & \cmark & \cmark
& \cmark & \cmark & \cmark & \cmark & \cmark
& \cmark & \cmark & \cmark & \cmark
& \cmark & \cmark & \cmark
& \cmark & \cmark & \cmark & \cmark \\
& RE2 & Auto
& \cmark & \cmark & \cmark
& \cmark & \cmark & \cmark & \cmark & \cmark & \cmark & \cmark & \cmark & \cmark
& \cmark & \xmark & \cmark & \xmark & \cmark & \xmark & \cmark & \xmark & \xmark & \xmark
& \cmark & \cmark
& \cmark & \cmark & \cmark & \cmark & \cmark
& \cmark
& \cmark & \cmark & -- & \cmark & \cmark & \cmark & --
& \cmark & \cmark & \cmark & \xmark & \cmark & \xmark
& \cmark & \xmark
& \cmark & \cmark & \xmark & \cmark & \xmark
& \cmark & \xmark & \cmark & \cmark & \xmark
& \xmark & \xmark & \xmark & \xmark
& \xmark & \xmark & \xmark
& \xmark & \xmark & -- & \xmark \\
\rowcolor{gray!10} & Oniguruma & BT
& \cmark & \cmark & \cmark
& \cmark & \cmark & \cmark & \cmark & \cmark & \cmark & \cmark & \cmark & \cmark
& \cmark & \xmark & \cmark & \cmark & \cmark & \cmark & \cmark & \cmark & \cmark & \cmark
& \cmark & \cmark
& \cmark & \cmark & \cmark & \cmark & \cmark
& \cmark
& \cmark & \cmark & \cmark & \cmark & \cmark & \cmark & \cmark
& \cmark & \cmark & \cmark & \cmark & \cmark & \cmark
& \cmark & \cmark
& \cmark & \cmark & \xmark & \cmark & \cmark
& \cmark & \xmark & \cmark & \cmark & \cmark
& \cmark & \cmark & \cmark & \cmark
& \cmark & \cmark & \cmark
& \cmark & \cmark & \qmark & \cmark \\
& Onigmo & BT
& \cmark & \cmark & \xmark
& \cmark & \cmark & \cmark & \cmark & \cmark & \cmark & \cmark & \cmark & \cmark
& \cmark & \xmark & \cmark & \xmark & \cmark & \xmark & \cmark & \xmark & \cmark & \cmark
& \cmark & \cmark
& \cmark & \cmark & \cmark & \cmark & \cmark
& \cmark
& \cmark & \cmark & \cmark & \cmark & \cmark & \cmark & \cmark
& \cmark & \cmark & \cmark & \cmark & \cmark & \cmark
& \cmark & \cmark
& \cmark & \cmark & \xmark & \cmark & \cmark
& \cmark & \xmark & \cmark & \cmark & \cmark
& \cmark & \cmark & \cmark & \cmark
& \cmark & \cmark & \cmark
& \cmark & \cmark & \xmark & \xmark \\
\rowcolor{gray!10} & Hyperscan & Hyb
& \cmark & \cmark & \cmark
& \cmark & \cmark & \cmark & \cmark & \cmark & \cmark & \cmark & \cmark & \cmark
& \cmark & \xmark & \cmark & \cmark & \cmark & \cmark & \cmark & \xmark & \xmark & \xmark
& \cmark$^\dagger$ & \cmark$^\dagger$
& \cmark & \cmark & \cmark & \cmark & \cmark
& \cmark
& \cmark & \qmark & \xmark & \cmark & \cmark & \cmark & \xmark
& \cmark & \cmark & \cmark & \cmark & \cmark & \xmark
& \qmark & \xmark
& \qmark & \cmark & \xmark & \qmark & \cmark
& \cmark & \xmark & \cmark & \cmark & \cmark
& \xmark & \xmark & \xmark & \xmark
& \xmark & \xmark & \xmark
& \xmark & \xmark & \xmark & \xmark \\
& TRE & Auto
& \cmark & \cmark & \xmark
& \cmark & \cmark & \cmark & \cmark & \cmark & \cmark & \cmark & \cmark & \cmark
& \cmark & \xmark & \cmark & \xmark & \cmark & \xmark & \cmark & \xmark & \xmark & \xmark
& \xmark & \xmark
& \cmark & \cmark & \cmark & \cmark & \xmark
& \cmark
& \cmark & \cmark & \xmark & \cmark & \cmark & \cmark & \xmark
& \cmark & \cmark & \cmark & \cmark & \cmark & \xmark
& \cmark & \xmark
& \cmark & \cmark & \xmark & \xmark & \cmark
& \qmark$^\ddagger$ & \xmark & \qmark$^\ddagger$ & \qmark$^\ddagger$ & \xmark
& \xmark & \xmark & \xmark & \xmark
& \cmark & \xmark & \xmark
& \xmark & \xmark & \xmark & \xmark \\
\rowcolor{gray!10} & mrab-regex & BT
& \cmark & \cmark & \cmark
& \cmark & \cmark & \cmark & \cmark & \cmark & \cmark & \cmark & \cmark & \cmark
& \cmark & \xmark & \cmark & \xmark & \cmark & \xmark & \cmark & \xmark & \cmark & \cmark
& \cmark & \cmark
& \cmark & \cmark & \cmark & \cmark & \cmark
& \cmark
& \cmark & \cmark & \cmark & \cmark & \cmark & \cmark & \cmark
& \cmark & \cmark & \cmark & \cmark & \cmark & \cmark
& \cmark & \cmark
& \cmark & \cmark & \cmark & \cmark & \cmark
& \cmark & \xmark & \cmark & \cmark & \cmark
& \cmark & \cmark & \cmark & \cmark
& \cmark & \cmark & \cmark
& \cmark & \cmark & \cmark & \xmark \\
& regexp2 & BT
& \cmark & \cmark & \xmark
& \cmark & \cmark & \cmark & \cmark & \cmark & \cmark & \cmark & \cmark & \cmark
& \cmark & \xmark & \cmark & \xmark & \cmark & \xmark & \cmark & \xmark & \xmark & \xmark
& \cmark & \cmark
& \cmark & \cmark & \cmark & \cmark$^\dagger$ & \cmark$^\dagger$
& \cmark
& \cmark & \cmark & \xmark & \cmark & \cmark & \cmark & \cmark
& \cmark & \cmark & \cmark & \cmark & \cmark & \cmark
& \cmark & \xmark
& \cmark & \cmark & \xmark & \cmark & \cmark
& \cmark & \xmark & \cmark & \cmark & \cmark
& \cmark & \cmark & \cmark & \cmark
& \cmark & \xmark & \cmark
& \xmark & \cmark & \xmark & \xmark \\
\rowcolor{gray!10} & fancy-regex & Hyb
& \cmark & \cmark & \xmark
& \cmark & \cmark & \cmark & \cmark & \cmark & \cmark & \cmark & \cmark & \cmark
& \cmark & \xmark & \cmark & \cmark & \cmark & \cmark & \cmark & \xmark & \xmark & \xmark
& \cmark & \cmark
& \cmark & \cmark & \cmark & \cmark & \cmark
& \cmark
& \cmark & \cmark & \cmark & \cmark & \cmark & \cmark & \cmark
& \cmark & \cmark & \cmark & \cmark & \cmark & \cmark
& \cmark & \cmark
& \cmark & \cmark & \xmark & \cmark & \cmark
& \cmark & \xmark & \cmark & \cmark & \cmark
& \cmark & \cmark & \cmark & \cmark
& \cmark & \cmark & \cmark
& \qmark & \cmark & \qmark & \xmark \\
& sregex & Auto
& \cmark & \cmark & \xmark
& \cmark & \cmark & \cmark & \cmark & \cmark & \cmark & \cmark & \cmark & \cmark
& \cmark & \xmark & \cmark & \cmark & \cmark & \cmark & \cmark & \xmark & \xmark & \xmark
& \xmark & \xmark
& \cmark & \cmark & \cmark & \xmark & \xmark
& \cmark
& \cmark & \cmark & -- & \cmark & \cmark & \cmark & --
& \cmark & \cmark & \cmark & \cmark & \cmark & \xmark
& \cmark & \xmark
& \cmark & \cmark & \xmark & \xmark & \xmark
& \qmark$^\ddagger$ & \xmark & \qmark$^\ddagger$ & \qmark$^\ddagger$ & \xmark
& \xmark & \xmark & \xmark & \xmark
& \xmark & \xmark & \xmark
& \xmark & \xmark & -- & \xmark \\
\rowcolor{gray!10} & tiny-regex-c & BT
& \cmark & \cmark & \xmark
& \xmark & \xmark & \xmark & \xmark & \cmark & \cmark & \cmark & \xmark & \xmark
& \cmark & \xmark & \cmark & \xmark & \cmark & \xmark & \cmark & \xmark & \xmark & \xmark
& \xmark & \xmark
& \cmark & \qmark & \cmark & \xmark & \xmark
& \xmark
& \cmark & \xmark & \xmark & \xmark & \xmark & \xmark & \xmark
& \cmark & \cmark & \xmark & \xmark & \xmark & \xmark
& \xmark & \xmark
& \xmark & \xmark & \xmark & \xmark & \xmark
& \xmark & \xmark & \xmark & \xmark & \xmark
& \xmark & \xmark & \xmark & \xmark
& \xmark & \xmark & \xmark
& \xmark & \xmark & \xmark & \xmark \\
\bottomrule
\end{tabular}
\begin{tablenotes}
\tiny
\item \cmark=Yes, \xmark=No, {\qmark}=Partial (\ie implementation is incomplete, buggy, or feature deliberately ignored), --=N/A (non-backtracking engine cannot support).
\item $^\dagger$This feature is not enabled by default and requires explicit opt-in flags to be passed to the regex engine.
$^\ddagger$The same functionality can be achieved via another construct/API.
\item $^a$Go's \texttt{regexp} and Rust's \texttt{regex} crate are based on RE2's design and syntax.
$^b$PHP's \texttt{preg\_*} functions wrap PCRE2.
$^c$Ruby's \texttt{Regexp} is based on Onigmo, but Ruby maintains a separate fork in its source tree.
\end{tablenotes}

%% file: tables/table_2.tex
\begin{tabular}{@{}lcccccccccccc@{}}
\toprule
\textbf{Engine}
& \rotatebox{90}{\textbf{Unit}}
& \rotatebox{90}{\textbf{Regression}}
& \rotatebox{90}{\textbf{Fuzzing}}
& \rotatebox{90}{\textbf{Differential}}
& \rotatebox{90}{\textbf{Property-Based}}
& \rotatebox{90}{\textbf{Performance}}
& \rotatebox{90}{\textbf{Memory Safety}}
& \rotatebox{90}{\textbf{Boundary}}
& \rotatebox{90}{\textbf{Integration}}
& \rotatebox{90}{\textbf{Exception Safety}}
& \rotatebox{90}{\textbf{Concurrency}}
& \rotatebox{90}{\textbf{Negative}} \\
\midrule
\multicolumn{13}{l}{\textit{First-Party (Language Standard Library)}} \\
\midrule
Python \texttt{re}       & \cmark & \cmark & \xmark & \xmark & \xmark & \cmark & \xmark & \cmark & \cmark & \xmark & \cmark & \cmark \\
\rowcolor{gray!10} Java \texttt{j.u.regex}  & \cmark & \cmark & \xmark & \xmark & \xmark & \cmark & \xmark & \cmark & \cmark & \xmark & \cmark & \cmark \\
JS/TS \texttt{Irregexp}  & \cmark & \cmark & \cmark & \xmark & \cmark & \cmark & \xmark & \cmark & \cmark & \xmark & \cmark & \cmark \\
\rowcolor{gray!10} C++ \texttt{<regex>}     & \cmark & \cmark & \xmark & \cmark & \xmark & \cmark & \xmark & \cmark & \cmark & \cmark & \xmark & \cmark \\
Go \texttt{regexp}       & \cmark & \cmark & \xmark & \cmark & \xmark & \cmark & \xmark & \cmark & \cmark & \xmark & \xmark & \cmark \\
\rowcolor{gray!10} C\# \texttt{S.T.Regex}   & \cmark & \cmark & \xmark & \cmark & \xmark & \cmark & \xmark & \cmark & \cmark & \xmark & \xmark & \cmark \\
Rust \texttt{regex}      & \cmark & \cmark & \cmark & \cmark & \cmark & \cmark & \xmark & \cmark & \cmark & \xmark & \xmark & \cmark \\
\rowcolor{gray!10} PHP \texttt{preg\_*}     & \cmark & \cmark & \cmark & \xmark & \xmark & \cmark & \xmark & \cmark & \cmark & \xmark & \xmark & \cmark \\
Ruby \texttt{Regexp}     & \cmark & \cmark & \xmark & \xmark & \xmark & \cmark & \cmark & \cmark & \cmark & \xmark & \cmark & \cmark \\
\rowcolor{gray!10} Shell \texttt{glibc}     & \cmark & \cmark & \xmark & \xmark & \xmark & \cmark & \cmark & \cmark & \cmark & \xmark & \xmark & \cmark \\
\midrule
\multicolumn{13}{l}{\textit{Third-Party Libraries}} \\
\midrule
PCRE                     & \cmark & \cmark & \xmark & \cmark & \xmark & \cmark & \cmark & \cmark & \cmark & \xmark & \xmark & \cmark \\
\rowcolor{gray!10} PCRE2                    & \cmark & \cmark & \cmark & \cmark & \xmark & \cmark & \cmark & \cmark & \cmark & \xmark & \xmark & \cmark \\
RE2                      & \cmark & \cmark & \cmark & \cmark & \cmark & \cmark & \cmark & \cmark & \cmark & \xmark & \cmark & \cmark \\
\rowcolor{gray!10} Oniguruma                & \cmark & \cmark & \cmark & \xmark & \xmark & \cmark & \cmark & \cmark & \cmark & \xmark & \xmark & \cmark \\
Onigmo                   & \cmark & \cmark & \xmark & \xmark & \xmark & \cmark & \xmark & \cmark & \cmark & \xmark & \xmark & \cmark \\
\rowcolor{gray!10} Hyperscan                & \cmark & \cmark & \xmark & \cmark & \xmark & \cmark & \xmark & \cmark & \cmark & \xmark & \xmark & \cmark \\
TRE                      & \cmark & \cmark & \cmark & \xmark & \xmark & \cmark & \cmark & \cmark & \cmark & \xmark & \xmark & \cmark \\
\rowcolor{gray!10} mrab-regex               & \cmark & \cmark & \xmark & \xmark & \xmark & \cmark & \xmark & \cmark & \cmark & \xmark & \xmark & \cmark \\
regexp2                  & \cmark & \cmark & \cmark & \cmark & \xmark & \cmark & \xmark & \cmark & \cmark & \xmark & \xmark & \cmark \\
\rowcolor{gray!10} fancy-regex              & \cmark & \cmark & \cmark & \cmark & \cmark & \cmark & \xmark & \cmark & \cmark & \xmark & \xmark & \cmark \\
sregex                   & \cmark & \cmark & \xmark & \cmark & \xmark & \cmark & \cmark & \cmark & \cmark & \xmark & \xmark & \cmark \\
\rowcolor{gray!10} tiny-regex-c             & \cmark & \cmark & \xmark & \cmark & \cmark & \cmark & \xmark & \cmark & \cmark & \xmark & \xmark & \cmark \\
\bottomrule
\end{tabular}

%% file: tables/table_3.tex
\begin{tabular}{@{}llll@{}}
\toprule
\textbf{Category} & \textbf{Relation} & \textbf{Equivalence} & \textbf{Assertion} \\
\midrule
\multirow{4}{*}{Alternation}
  & Associativity & $r_1|(r_2|r_3) \equiv (r_1|r_2)|r_3$ & match \\
  & Commutativity & $r_1|r_2 \equiv r_2|r_1$ & match \\
  & Idempotence & $r \equiv r|r$ & match \\
  & Zero Identity & $r|\emptyset \equiv r$ & match \\
\midrule
\multirow{3}{*}{Concatenation}
  & Associativity & $(r_1r_2)r_3 \equiv r_1(r_2r_3)$ & match \\
  & One Identity & $r \cdot \varepsilon \equiv r$ & match \\
  & Zero Identity & $r \cdot \emptyset \equiv \emptyset$ & no match \\
\midrule
\multirow{2}{*}{Distributivity}
  & Left & $r_1(r_2|r_3) \equiv r_1r_2|r_1r_3$ & match \\
  & Right & $(r_1|r_2)r_3 \equiv r_1r_3|r_2r_3$ & match \\
\midrule
\multirow{4}{*}{Kleene Star}
  & Left Unrolling & $r^* \equiv (\varepsilon | rr^*)$ & match \\
  & Right Unrolling & $r^* \equiv (\varepsilon | r^*r)$ & match \\
  & Collapse & $(r^*)^* \equiv r^*$ & match+span \\
  & Expand & $r^* \equiv (r^*)^*$ & match+span \\
\midrule
\multirow{3}{*}{Star Laws}
  & Sum-Star Left & $(r_1|r_2)^* \equiv r_1^*(r_2r_1^*)^*$ & match \\
  & Sum-Star Right & $(r_1|r_2)^* \equiv (r_1^*r_2)^*r_1^*$ & match \\
  & Product-Star & $(r_1r_2)^* \equiv \varepsilon | r_1(r_2r_1)^*r_2$ & match \\
\bottomrule
\end{tabular}